\documentclass[12 pt]{article}
\usepackage[utf8]{inputenc}
\usepackage[english]{babel}
\usepackage{authblk}
\usepackage{graphicx}
\usepackage{cite}
\usepackage{color}
\usepackage{gensymb}

\newcommand {\angstrom} {\AA} 
\def\um    {\ensuremath{\mu}m}
\newcommand {\nm} {{\rm nm}}

\title{ Hard X-ray multi-projection imaging for single-shot approaches }
\author[1,2,*]{P. Villanueva-Perez}
\author[1]{B.~Pedrini}
\author[1,3]{R.~Mokso}
\author[2]{P.~Vagovic}
\author[1]{V.~Guzenko}
\author[4]{S.~Leake}
\author[1]{P.~R.~Willmott}
\author[1]{C.~David}
\author[2,5,6]{H.~N.~Chapman}
\author[1,7]{M.~Stampanoni}
\affil[1]{Paul Scherrer Institut, Villigen, Switzerland}
\affil[2]{Center for Free-electron Laser Science (DESY), Hamburg, Germany}
\affil[3]{Max IV Laboratory, Lund University, Lund, Sweden}
\affil[4]{ESRF - The European synchrotron, Grenoble, France}
\affil[5]{University of Hamburg, Hamburg, Germany}
\affil[6]{Centre for Ultrafast Imaging, Hamburg, Germany}
\affil[7]{Institute for Biomedical Engineering, UZH/ETH Z\"urich, Z\"urich, Switzerland}
\affil[*]{Corresponding author: P.~Villanueva-Perez, pablo.villanueva-perez@cfel.de}

\date{}                    
\begin{document}
\maketitle

\begin{abstract}
  %
  %
  %
  %
  Obtaining 3D information from a single X-ray exposure at high-brilliance sources, such as X-ray free-electron lasers (XFELs)~\cite{Emma2010} or diffraction-limited storage rings~\cite{Eriksson2014}, allows the study of fast dynamical processes in their native environment.
  However, current X-ray 3D methodologies are either not compatible with single-shot approaches because they rely on multiple exposures, such as confocal microscopy~\cite{davidovits1969,Vincze2004} and tomography~\cite{Cormack1964,Hounsfield1973}; or they record a single projection per pulse~\cite{Raines2010} and are therefore restricted to approximately two-dimensional objects~\cite{Wei2011}.
  Here we propose and verify experimentally a novel imaging approach named X-ray multi-projection imaging (XMPI), which simultaneously acquires several projections without rotating the sample at significant tomographic angles. 
  When implemented at high-brilliance sources it can provide volumetric information using a single pulse. Moreover, XMPI at MHz repetition XFELs could allow a way to record 3D movies of deterministic or stochastic natural processes in the micrometer to nanometer resolution range, and at time scales from microseconds down to femtoseconds.
\end{abstract}

Since their discovery, hard X-rays have been crucial in natural sciences because of their penetration power and short wavelength, which allows high-resolution imaging of thick samples, even in native conditions.
Among the currently used X-ray imaging techniques, phase-contrast methods enhance the contrast sensitivity by exploiting the phase shift due to variations in the electron density rather than the intensity attenuation characteristic of radiographic approaches~\cite{Sakdinawat2010}. Coherent techniques, which exploit phase contrast, are regarded as the most suitable to achieve high-resolution~\cite{Chapman2010a}, in that they can address micrometer to nanometer scales. 
Because the high brilliance is the key parameter for coherent techniques, their advent coincided with the realization of third generation synchrotron light sources. 
Novel X-ray sources with orders of magnitude higher brilliance, such as diffraction limited storage rings~\cite{Eriksson2014} and X-ray free electron lasers (XFELs)~\cite{Emma2010,Ishikawa2012,Pellegrini2016}, enlarge the spectrum of coherent applications, especially addressing shorter timescales~\cite{Barty2010,Weckert2015}. 
XFELs in particular provide ultraintense femtosecond pulses which can image samples before inducing any radiation damage~\cite{Neutze2000}.
This concept, known as diffract before destroy, was demonstrated experimentally~\cite{Chapman2006a} by reconstructing an object from
an X-ray pulse but before it Coulomb explodes.
The resolution and contrast sensitivity are limited by the number of photons available in a single pulse, and not by the maximum tolerable dose which preserves a given resolution~\cite{Howells2009}, as is the case for continuous sample illumination.
As a consequence, any method that requires multiple exposures of the same sample, including three-dimensional (3D) techniques such as tomography~\cite{Cormack1964,Hounsfield1973} and confocal microscopy~\cite{davidovits1969}, or any scanning technique cannot be applied.
Thus, XFEL applications aiming at 3D structural information, which deliver high dose, either require imaging of several identical copies of the object~\cite{Ekeberg2015}, or are restricted to retrieving partial information from a single exposure~\cite{Wei2011}, as desired in ankylography~\cite{Raines2010}.

\begin{figure}[htbp!]
  \centering
  \includegraphics[width=0.95\textwidth]{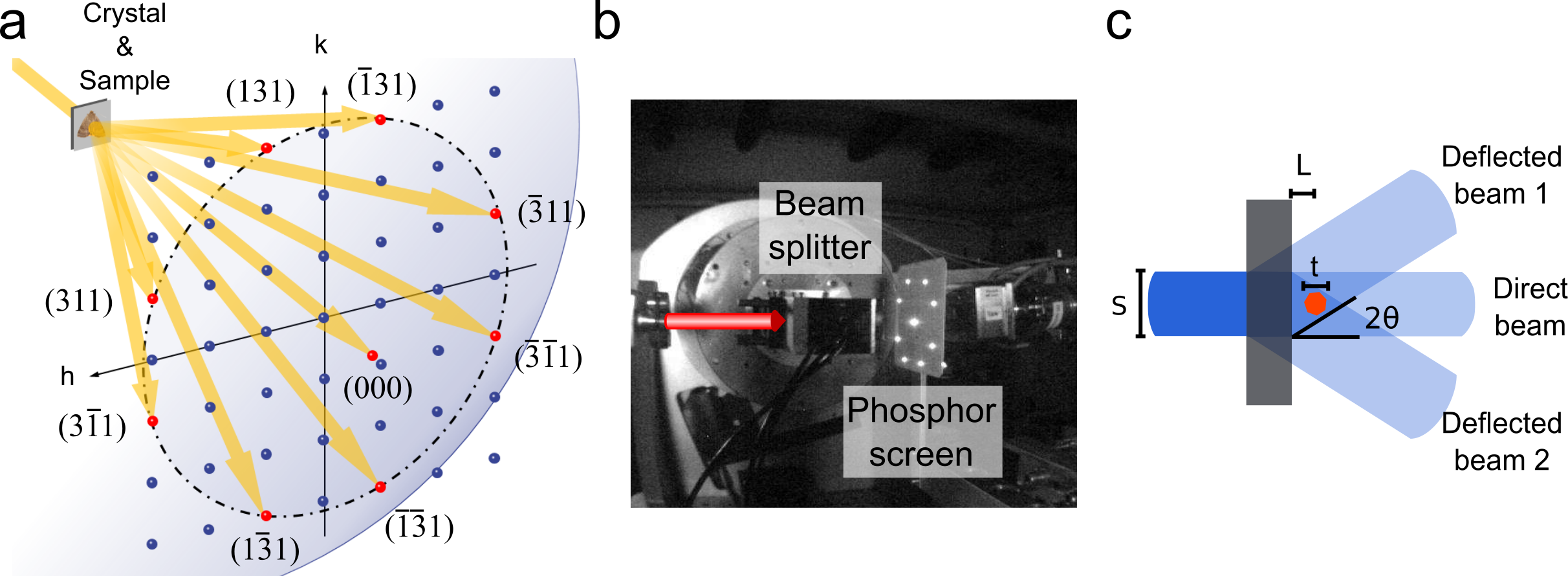}
  \caption{Beam splitter.
   (a) Illustration of the eight reflections in reciprocal space of the \{113\}-family of a face-centered cubic crystal. The dotted-dashed curve represents the intersection between the Ewald sphere and the l=1 plane. The sample is positioned downstream the crystal to be illuminated by all the generated beams.
  (b) Image of the direct beam and of the eight diffracted beams on a phosphor screen, generated from a single incoming beam traversing a 100~$\mu$m thick Si(001) crystal perpendicularly to the (001) surface.
  (c) Representation of the requirement to the maximum distance $L$ from the crystal beam splitter surface to the sample such that the latter is illuminated by both the direct and the diffracted beams. The relevant parameters are the diameter of the direct beam $S$, the transverse extension of the sample $t$ and the deflection angle $2\theta$ of the diffracted beam.
  \label{fig:splitter}}
\end{figure}

Here we propose a scheme christened X-ray multi-projection imaging (XMPI), which provides 3D structural information via multiple 2D projections at different tomographic angles acquired simultaneously from the same object.
The key component of XMPI is a beam splitter that generates a number of beams by Laue diffraction, which illuminate a sample simultaneously from different angles.
Each of these beams retains the corresponding projection information.
This idea was proposed in 1994 for the soft X-ray regime~\cite{Howells1994a} using a phase-grating splitter.
In the hard X-ray regime under consideration, however, suitable gratings are unrealistic;
for example, the grating pitch to achieve a deflection for the first diffracted order of $20^{\circ}$ for 4~keV photons would be 8.5~\angstrom, which is too small for presently-known manufacturing methods.
In contrast, Laue diffracted beams are much more suitable because the deflection angles reach tens of degrees, compatible with the requirement for true tomographic projections.
In a general case, the Laue condition can be achieved simultaneously for two different reflections by appropriately orienting the crystal~\cite{Schmidt2008}.
This number can be increased by exploiting symmetries of the crystal lattice, setting the X-ray energy to specific values, and positioning the crystal so that several reflections 
sit simultaneously on the Ewald sphere.
Figure~\ref{fig:splitter}(a) illustrates the generation of eight deflected beams by the \{113\}-family of Laue reflections of a face-centered-cubic crystal, such as diamond or silicon.
The incoming beam direction, defined by its momentum vector $\vec{k_0}$, is set parallel to a high-symmetry axis, corresponding to the (001)-reflection direction in the depicted example. 
All reflections related by a rotation around the symmetry axis, e.g. corresponding to \{113\}-family, form identical angles $\pi/2-\theta$ with respect to the incoming beam direction and share the same reflection plane spacing $d$. 
The Laue condition for the wavelength $\lambda$, 
\begin{equation}
  \label{eq:Bragg}
  \lambda=2d\sin(\theta)~,
\end{equation}
is then fulfilled simultaneously by all eight planes, yielding eight diffracted beams with a deflection angle of $2\theta$.
For a silicon crystal, the photon energy that sets the \{113\} planes in the Laue condition is $12.56$~keV (see Methods).
Figure~\ref{fig:splitter}(b) provides experimental evidence for the simultaneous generation of the eight beams described above. 
The experiment was performed at the Material Science beamline~\cite{Willmott2013} of the Swiss Light Source (SLS), using a silicon crystal with the aforementioned arrangement.
Figure~\ref{fig:splitter}(c) sketches the arrangement of the beam-splitter crystal and a sample positioned downstream the crystal in the overlap region of all eight diffracted beams.
To ensure the simultaneous illumination of a sample of size $t$ by all the beams, the incoming beam diameter $S$ and maximum distance from the sample center to the closest face of the crystal $L$ are constrained (see Methods). 


XMPI is a technique which can be applied to the near-field and far-field imaging regimes.
In this work, we demonstrate that the different projections of an object are retrieved for both regimes with resolutions around 17~\um~and 80~\nm, respectively.

\begin{figure}[htbp!]
  \centering
  \includegraphics[width=0.95\textwidth]{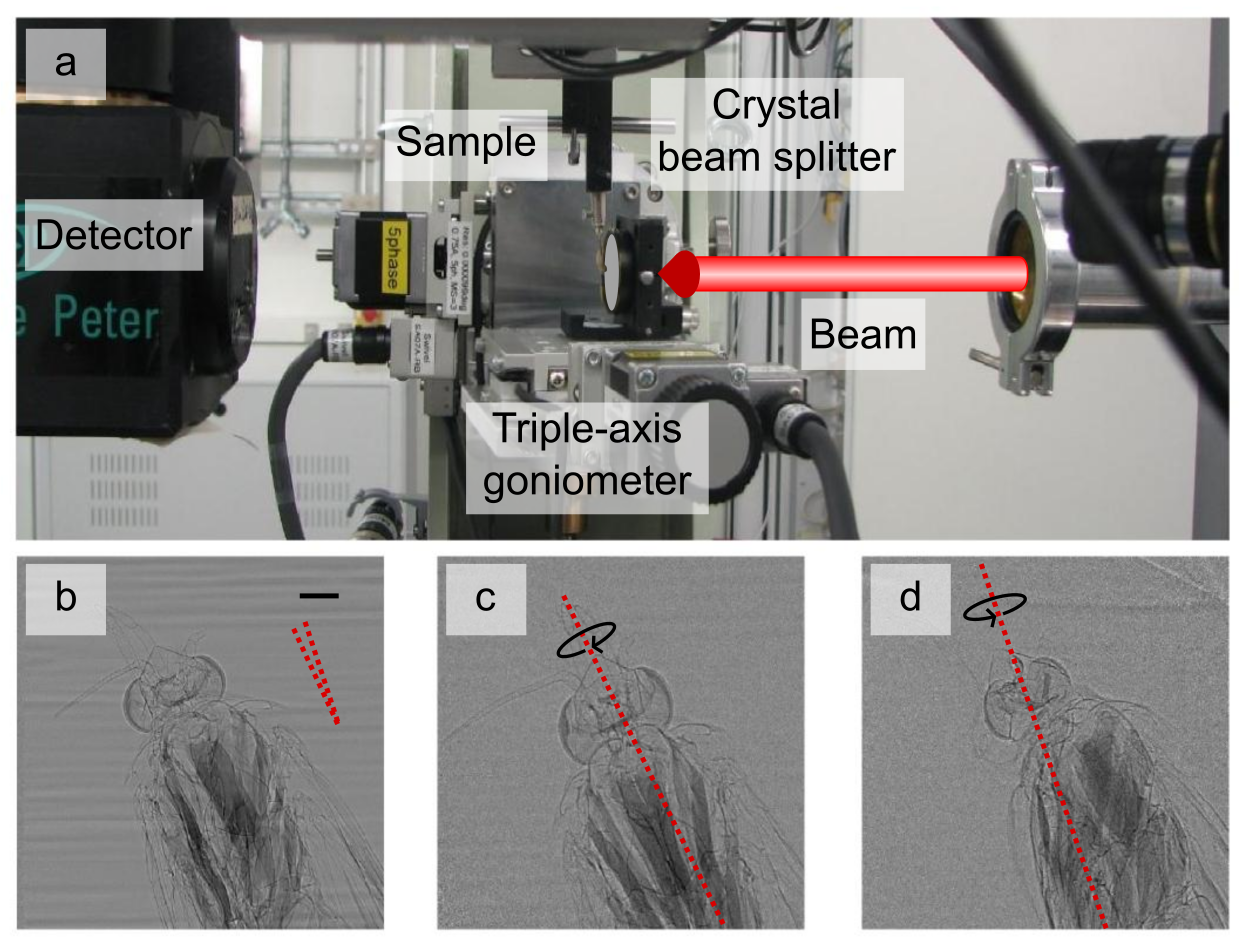}
  \caption{Near-field imaging experiment.
  (a) Experimental setup used at the TOMCAT beamline of the Swiss Light Source.
  (b-d) Phase contrast images in the near-field regime recorded  with the area detector placed in the horizontal plane at deflection angles of~(b) $0^{\circ}$ (direct beam direction),~(c)  $18.2^{\circ}$ (diffracted beam from the Si(111) reflection), and~(d) $35.1^{\circ}$ (diffracted beam for the Si(311) reflection), respectively.
  The detection plane was perpendicular to the direct beam.
  The rotation axes and rotation directions with respect to the projection in (a) are marked with dashed-red lines and black arrows.
  The scale bar in~(b) corresponds to 500~$\mu$m and the two red-dashed lines illustrate the angle between the rotation axes.\label{fig:near-field}}
\end{figure}

The near-field imaging experiment was carried out at the TOMCAT beamline at SLS~\cite{Stampanoni2006}.
Propagation-based phase-contrast imaging was performed (see Methods) using the setup depicted in Fig.~\ref{fig:near-field}(a). 
The collimated beam at 12.56~keV illuminated a Si(001) splitter, mounted on a triple-axis goniometer.
Due to geometrical limitations of the experimental setup, the crystal could not be oriented to hit simultaneously the eight reflections of the Si \{113\}-family (Fig.~\ref{fig:splitter}(a)), but only the Si(131) and the Si(111) reflections with deflections angles of $35.1^{\circ}$ and $18.2^{\circ}$, respectively.
A moth placed directly downstream of the splitter was illuminated simultaneously by the three beams.
Three near-field images shown in  Fig.~\ref{fig:near-field}(b-d) were recorded by translating the detector to intercept each of the three beams. 
The forward-direction image exhibits lower noise because of the higher intensity.
The image resolution of such images was estimated to be about $17~\mu$m by analyzing the edge profiles.
The rotation axes that relate the direct-beam projection (b) with the two deflected beam projections (c) and (d) form the expected angle of $11.9^{\circ}$.
The features of the moth head observed in the three images concur with being projections of the same object along the directions given by (001), (111), and ($\bar{1}\bar{3}1$).

\begin{figure}[htbp!]
  \centering
  \includegraphics[width=0.95\textwidth]{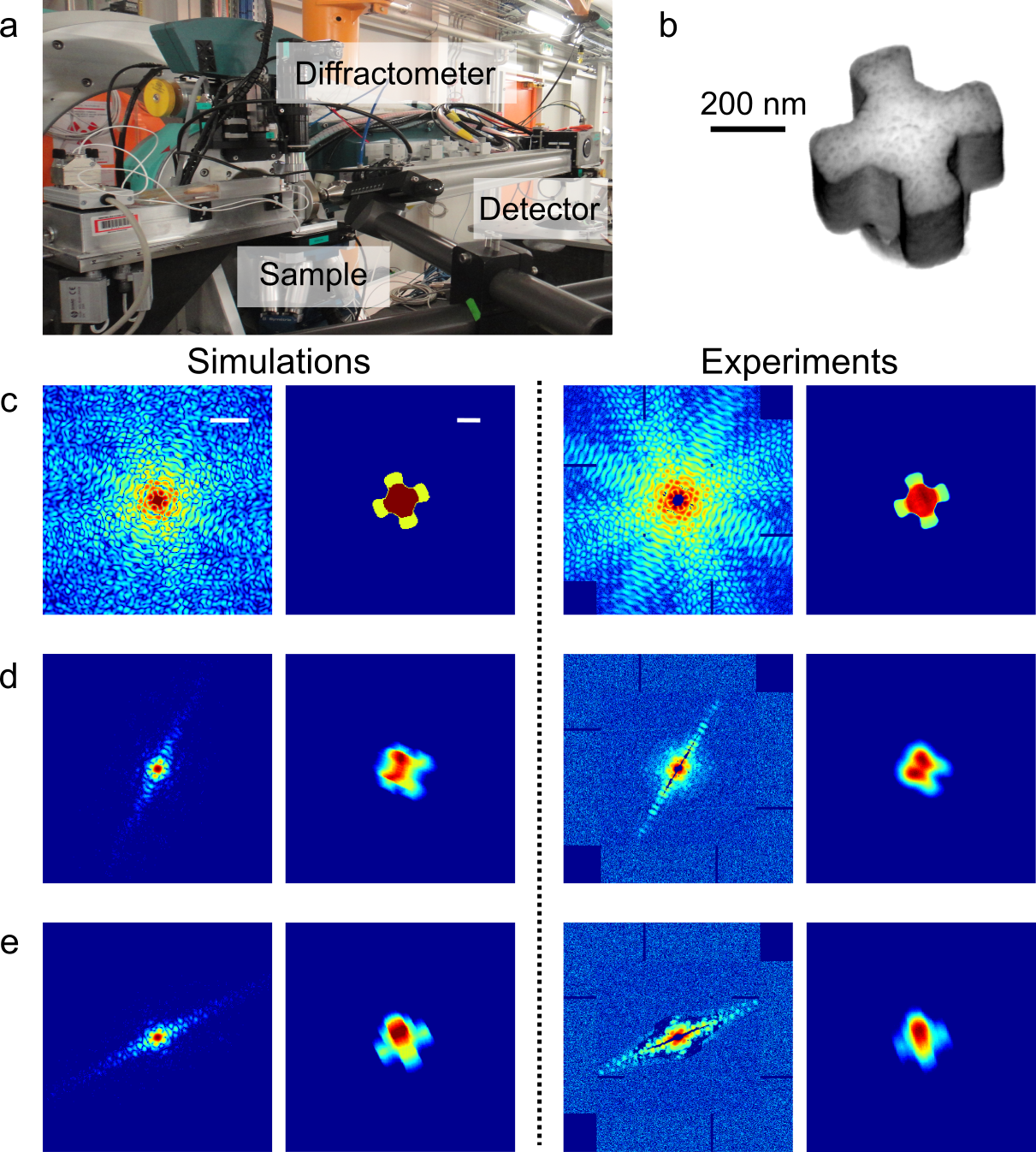}
  \caption{Far-field imaging experiment.
  (a) Experimental setup used at ID01 beamline of the ESRF synchrotron.
  (b) SEM image of the gold nanostructure sample.
  (c-e) Data related to the direct beam (c) and to the two accessible projections (d,e). From left to right: simulated diffraction pattern, corresponding simulated object projection, experimental diffraction pattern, corresponding CDI reconstruction.
  The scale bars in the diffraction patterns and in the reconstructions correspond to $2\times 10^{-2}$~nm$^{-1}$ and 200~nm, respectively.
  \label{fig:far-field}}
\end{figure}
\begin{figure}[htbp!]
  \centering
  \includegraphics[width=0.95\textwidth]{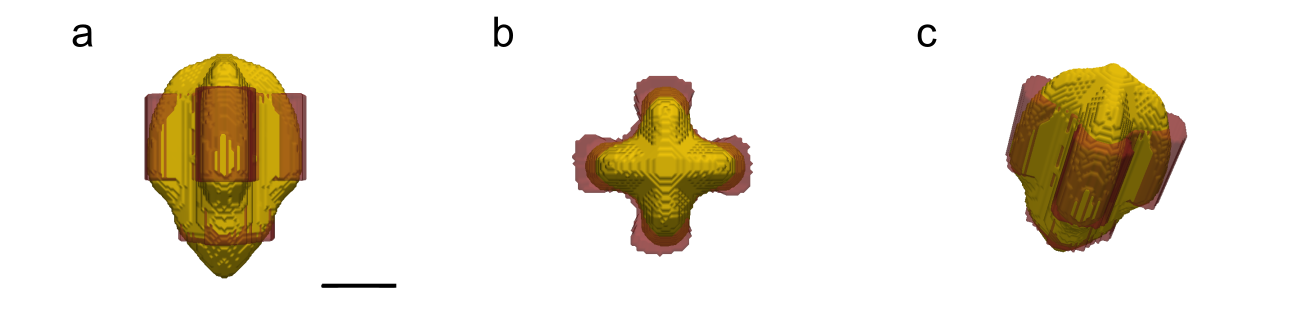}
  \caption{3D reconstructions. These panels depict the retrieved object (yellow) compared to the simulated phantom (semi-transparent red) projected along (a) a direction perpendicular to the direct beam, (b) the direct beam, and (c) an arbitrary direction. The scale bar in (a) corresponds to 200 nm.
  \label{fig:3d}}
\end{figure}

The far-field imaging experiment at 12.56~keV was performed at the ID01 beamline of the European Synchrotron Research Facility (ESRF)~\cite{Leake2017}. We performed coherent diffraction imaging (CDI)~\cite{Miao1999} (see Methods), a well-established technique at storage rings and XFELs~\cite{Chapman2010a}, using the setup shown in Fig.~\ref{fig:far-field}(a) (see Methods). 
A Si(001) splitter was mounted on a small hexapod to adjust the orientation.
The crystal was oriented such that the two \{113\}-family diffracted beams accessible in the ID01 diffractometer geometry were seen simultaneously on a pixel detector. 
A gold nanostructure, exhibiting non-trivial 3D features (Fig.~\ref{fig:far-field}(b)) and grown on a silicon nitride membrane (see Methods), was glued on the downstream surface of the crystal.
As the coherent flux was not sufficient, the beam was focused to a size about $S=1~\mu\rm{m}$
at the crystal surface with a numerical aperture which matched the Darwin width of the Si(131) reflection~\cite{Darwin1914,Darwin1914a}.
Unfortunately, $S$ did not allow the simultaneous illumination of the sample by the different beams as desired. However, this is not a limitation at sources with higher coherent flux such as XFELs and diffraction-limited synchrotrons.
The sample was then translated transversely to produce diffraction patterns on the detector positioned at 2.37~m distance.
The three recorded diffraction patterns are shown in the third column of panels (c-e) of Figure~\ref{fig:far-field}, along with corresponding simulations (first column) with same signal levels.
The experimental patterns from the diffracted beam clearly manifest larger background levels due to lower flux and backgrounds components enhanced by the crystal.
The CDI reconstructions from the experimental diffraction images, obtained by applying phase-retrieval algorithms (see Methods), and the simulated projections of the sample are shown in the fourth and second column of the same panels as above.
Their comparison confirms that the expected projections have been measured.
The resolution of the reconstructions, established using the phase-retrieval transfer-function criterion~\cite{Shapiro2005d}, was 18~nm for the direct beam projection and 77 and 85~nm for the two skew projections.
A 3D reconstruction of the object using the three measured projections is depicted in Fig.~\ref{fig:3d}. The reconstruction in yellow is compared to the simulated model in semi-transparent red. 


In conclusion, we have validated experimentally XMPI that relies on a single crystal as beam splitter to generate simultaneously tomographic projections from a single exposure of a sample to the X-rays.
XMPI circumvents rotating the sample as for tomography, and represents a clear improvement with respect to pseudo-3D single-shot methods.
We conceived XMPI as an imaging method for XFELs.
In the diffract-before-destroy approach, essential to achieve sub-micrometer resolution from weakly scattering, non-reproducible objects, XMPI paves the way to 3D object reconstructions.
Other applications appear however to be meaningful.
If XFELs that offer pulse trains at MHz repetition rates, such as the European XFEL or the Linac Coherent Light Source after the planned upgrade, are operated at fluences below the sample damage threshold, XMPI may enable to track 3D structural dynamics of stochastic and deterministic~\cite{Schropp2015} processes at the submicrosecond time scale. 
At synchrotron facilities, XMPI may find applications in the case that a sample cannot be rotated due to the complexity of the sample environment.
Furthermore, at diffraction-limited sources, such as MAX-IV and future ones, the time resolution for structural dynamics investigations may be reduced well below the millisecond regime.
We therefore anticipate that dedicated XMPI instruments may be realized at operational and future hard X-ray user facilities.


\section*{Methods}
\label{sec:methods}

\subsection*{Crystal beam splitters}
\label{sec:crystals}
We propose the use of face-centered-cubic crystals as beam splitters due to their high degree of symmetry, i.e. they can potentially generate simultaneously multiple deflected beams. Specifically, we focus on diamond and silicon crystals. Diamond is a good candidate for X-ray free-electron lasers optics due to its low X-ray absorption, damage threshold, and good heat conductivity. On the other hand, silicon is a crystal which can be produced inexpensively, and with high purity, low strain, and no defects. Table~\ref{tab:cryst} reports the deflection angles for a few allowed silicon and diamond reflection families whose corresponding wavelengths are in the hard X-ray regime and with practical deflection angles.
The energy to set a family of planes in Bragg condition is given by
\begin{equation}
E=\frac{hc}{2d\sin\theta}~,
\end{equation}
where $E$ is the energy, h is the Planck constant, and c is the speed of light in vacuum.

\begin{table}[h]
\centering
  	\begin{tabular}{ccccc}
	Refl. & Symm. direction & Refl. mult. & $E$ & Deflection angle $2\theta$\\
	\hline
		Si(131) & (001) & 8 & 12.56 keV & 35.1$^{\circ}$\\
		Si(331) & (001) & 8 & 13.70 keV & 48.2$^{\circ}$\\
		~C($\bar{3}11$) & (011) & 6 & 13.52 keV & 50.5$^{\circ}$\\
		~C($1\bar{1}\bar{1}$) & (111) & 3 & 9.03 keV & 38.9$^{\circ}$\\
		~C($\bar{1}1\bar{3}$) & (111) & 6 & 11.04 keV & 63.0$^{\circ}$
	\end{tabular}
	\caption{
	Bragg-reflection families of diamond and silicon cubic crystal structures suitable
	for multi-beam generation in the photon energy range ($E$) between
        $2$ and $14$~keV, with a deflection angle between  $20^{\circ}$ to $65^{\circ}$.\label{tab:cryst}}
\end{table}

\subsection*{Simultaneous illumination geometry}
\label{sec:simultaneous}
In order to image the sample simultaneously by all the generated beams, two conditions have to be satisfied. First, we impose a geometric condition, which requires that all the beams fully illuminate the sample. Therefore, the sample has to be positioned downstream the crystal at a position $L\leq L_g$, where the maximum geometrically allowed distance $L_g$ for a thin crystal is constrained by the beam diameter $S$, the maximum transverse dimension of the sample $t$, and the deflection angle $2\theta$:
\begin{equation}
  \label{eq:Lg}
  Lg\leq \frac{1}{sin(2\theta)}\left(\frac{S\cos(2\theta)-t}{2} \right)~.
\end{equation}
Second, the direct and the deflected beams have different optical paths, thus they do not illuminate exactly at the same time the sample. Thus given a maximum tolerable time delay ($\Delta t$) which ensures the immutability of the sample, the maximum tolerable distance between the crystal and the sample is given by
\begin{equation}
  \label{eq:Lt}
  L_t=\frac{c\Delta t}{\frac{1}{cos(2\theta)}-1}~,
\end{equation}
where $c$ is the speed of light.
At XFELs, the maximum tolerable time delay between deflected and the direct beam is constrained by time interval between the arrival of the imaging pulse and the observation of radiation damage. For example, for a biological sample like a Lysozyme crystal imaged by $3\times 10^{12}$ photons at 12~keV focused on a $100\times100$~nm$^2$ the aforementioned interval is below 10~fs~\cite{Neutze2000} and the maximum distance is of the order of $10$~$\mu$m.
Finally, the sample to crystal distance $L$ is chosen to be smaller than the minimum of the geometrical $L_g$ and temporal $L_t$ constraints.

\subsection*{Propagation-based phase contrast}
\label{sec:phase-contrast}
Phase-contrast imaging techniques exploit the phase change of the exit wavefront after transmitting through a sample rather than the change of transmission due to absorption.
Such techniques are specially useful to distinguish between two materials with similar transmission or transparent materials to the probing radiation. The high penetration power of hard X-rays, specially for low-Z materials, makes them a perfect radiation type to exploit phase-contrast techniques.
In the context of this work, we exploit propagation-based phase contrast, i.e. we use the free-space propagation to observe the phase change of the exit wavefront as intensity variations in the recorded images~\cite{Nugent2010}.
The phase information is retrieved from the intensity variations using phase-retrieval algorithms. The main solutions to this inversion problem in the near-field regime linearize either the transmissivity of the sample as exploited by the contrast-transfer function approach (CTF)~\cite{Guigay1977} or the propagator as transport-of-intensity equations (TIE)~\cite{Teague1983} do. As the presented images are acquired at propagation distances of the order of the depth of focus of the used microscope, we can exploit TIE algorithms, such as that presented in Ref.~\cite{Paganin2002a}.

\subsection*{TOMCAT setup and data collection}
\label{sec:TOMCAT}
The phase-contrast near-field experiments were performed at the TOMCAT beamline of the Swiss Light Source. The X-rays provided by a bending magnet source were monochromatized by a multilayer monochromator to 12.6~keV with approximately a 2~\% bandwidth.
The natural divergence of the bending magnet X-ray beam at TOMCAT is about 2 mrad in the horizontal and 0.6 mrad in the vertical direction, which is larger than the Darwin width of the used crystal. 
A 100~$\mu$m thick Si(001) crystal was illuminated by a  $10\times 6$~$\rm{mm}^2$ beam after conforming it with three sets of slits. The crystal was mounted on a triple-axis goniometer to generate simultaneously several reflections. Behind the crystal a moth was positioned to be illuminated by the different generated beams. At 10 cm from the moth the detector was positioned to record the different phase-contrast images.
For each of the images the detector was translated in a plane perpendicular to the incoming beam.
The detector was an X-ray 1:1 (Optique Peter) microscope with a high efficiency scintillator, which converts X-rays to optical photons. The camera used was a pco.edge 4.2 CMOS detector with a pixel size of 6.5~$\mu$m and $2048\times2048$ pixels. Each of the moth projections was acquired with $4\times 10^{11}$~ph/$\rm{mm}^2$ on the crystal beam splitter~\cite{Lovric2016}. Around 63~\% of the fluence on the crystal beam splitter contributed to form the direct-beam image, while the contribution to the silicon (111) and (131) images was of 4~\% and 2~\%, respectively.

\subsection*{Coherent diffraction imaging}
\label{sec:cdi}
Coherent diffraction imaging (CDI)~\cite{Miao1999} is a lensless technique. The sample is illuminated by plane waves, and the diffraction patterns are recorded in the far-field regime. CDI provides a reconstruction of the complex X-ray transmissivity of the sample by means of phase-retrieval procedures based on iterative transform algorithms~\cite{Gerchberg1972,Fienup1980}. The achievable resolution is given by the largest diffraction angle at which the intensity of the diffraction pattern exhibits sufficient signal-to-noise in order to reliably reconstruct the phase~\cite{Nugent2010}, so that the ultimate resolution limit is set by the wavelength of the incident radiation.
The reconstruction presented in Fig.~\ref{fig:far-field}(c), (d), and (e) are obtained after averaging 20 reconstructions, where each of them was obtained after 2800 iterations of shrink-wrap algorithm~\cite{Marchesini2003a} and 1200 iteration of hybrid input-output~\cite{Fienup1980} with $\beta=0.9$ combined with an error reduction algorithm~\cite{Gerchberg1972}.

\subsection*{3D reconstruction}
\label{sec:3d}
The 3D reconstruction was retrieved by using the filtered backprojection algorithm. As the experimental data recorded at ID01 was limited to only three projections, we have applied symmetry constraints. First, we applied a four-fold symmetry constraint around the beam direction, i.e. perpendicular to the view in Fig.~\ref{fig:far-field}(c). Second, we applied a mirror symmetry constraint around a mirror plane defined perpendicular to the projection in Fig.~\ref{fig:far-field}(c). Once the 3D model was reconstructed, a histogram constraint vetoing the outlaiers was applied. The 3D data visualization has been obtained using ParaView software~\cite{Ahrens2005}.

\subsection*{ID01 setup and data collection}
\label{sec:id01}
The CDI experiments were performed at the ID01 beamline at ESRF.
The beam was monochromatized by a double crystal monochromator at 12.56~keV with a $\sim 4\times 10^{-4}$ bandwidth as required to efficiently generate eight beams by a silicon crystal (Table~\ref{tab:cryst}).
The coherent portion of the beam was focused by compound refractive lenses to a focal spot of $1~\mu\rm{m}^2$ on the sample. The sample was composed of a 100~$\mu$m thick Si(001) crystal and 500 nm gold nanostructures attached to it. The sample was mounted on a hexapod stage capable to precisely orient the sample to generate the eight deflected beams. The intensity of the deflected beams was 10~\% less intense than the transmitted beam. Once the crystal was aligned, a SmarAct 3D piezo system was used to position precisely the gold nanostructures to be illuminated by the deflected beams. As the beam was tightly focused around the sample and the sample was not exactly positioned behind the crystal, we translated around 50~$\mu\rm{m}$ the sample from the position in the direct beam to the position in each of the diffracted beams. The diffractometer was positioned on each of the three accessible beams to record the diffraction patterns. The detector used was a Maxipix with 55~$\mu$m pixel size and $512\times 512$ pixels. The Maxipix was positioned 2.37~m from the sample on the diffractometer arm and a vacuum pipe with a beamstop was installed between the sample and the detector to reduce the air scattering. The direct-beam image was recorded with a fluence of $1.1\times 10^{10}$~ph/$\mu\rm{m}^2$ illuminating the crystal beam splitter, but only 38~\% of that fluence illuminated the sample. The deflected-beam images were recorded for both reflections with a fluence of $3.0\times 10^{11}$~ph/$\mu\rm{m}^2$ with an efficiency of 4~\%. 

\subsection*{Sample preparation for CDI}
\label{sec:sample-preparation}
Gold nanoparticles for CDI experiments were manufactured on 250~nm thick $\rm{Si}_3\rm{N}_4$ membranes by means of two steps: electron-beam lithography and electroplating~\cite{Gorelick2010,Guzenko2014a}. First, a metal stack of Cr/Au/Cr (5nm/10nm/5nm) was evaporated on a silicon nitride membrane. Subsequently, a negative tone e-beam resist HSQ (FOX16, Dow Corning, 1:1 dilution with MIBK) was spin-coated at 3000~rpm, resulting in a film thickness of about 250~nm. In the first e-beam lithography step, an array of micro-rings were exposed using Vistec EBPG5000Plus direct writing e-beam lithography system, operated at 100~kV accelerating voltage. After development of the exposed HSQ in NaOH buffered solution (MICROPOSIT 351, Rohm and Haas) and rinsing in deionized water, silicon dioxide discs with inner/outer diameter of 250~nm/4~$\mu$m, respectively, and the distance between the neighboring particles of 100~$\mu$m were defined at specified locations. For the following e-beam exposure, approximately 900~nm thick layer of positive tone resist (PMMA 950k, 8\% in anisole, MicroChem Corp.) was spin-coated on the membrane at 4000~rpm and baked out at $175~^{\degree}$C on a hotplate. By performing the second overlay exposure and developing the samples in IPA:DI $\rm{H_2O}$ (7:3) solution, structures with various manifold rotational symmetry, fitting into a circle of 500~nm diameter, were created exactly above the $\rm{SiO}_2$ discs. This way, molds for electroplating gold nanoparticles with well-resolved 3D shape control were defined. After a short $\rm{Cl_2}$-based plasma etching step (required to remove the upper Cr layer), the mold was filled with gold during the electroplating step to a height of 500~nm. After removing the PMMA layer and subsequently HSQ discs in acetone and BOE (buffered oxide etch), respectively, individual nanoparticles with complex 3D shape along the rotational symmetry axis anchored on the $\rm{Si_3N_4}$ membrane were fabricated, an example is shown in Fig.~\ref{fig:far-field}(b).

\section*{Acknowledgments}
\label{sec:acknowledgements}
We are grateful to P. B\"osecke, T. Celcer, Y. Chushkin, H. Djazouli, M. Gordan, M. Lange, D. Meister, P. Oberta, and T.~Schulli, for their support in setting up the infrastructure for the experiments, and M. Guizar-Sicairos, A. Irastorza-Landa, G. Lovric, T. White, O. Yefanov, the TOMCAT group, and the coherent X-ray imaging group at CFEL for fruitful discussions and comments.
Part of this work has been supported by the ERC grant ERC-2012-StG 310005-PhaseX. We also acknowledge funding within the Röntgen-Angström-Cluster by the German Ministry for Education and Research (BMBF) under grant number 05K18XXA and by the Swedish Research Council under grant number 2017-06719.

\section*{Author contributions}
\label{sec:contributions}
P.V.-P., B.P., R.M., and M.S. initially conceived the research project, proposed the experiments, and developed the main concept of XMPI. P.V.-P. and P.V. devised the beam splitter concept and imaging implementations for XFELs and diffraction limited synchrotrons.
P.V.-P., B.P., R.M., and P.R.W. demonstrated experimentally the beam splitter at the MS beamline.
P.V.-P., B.P., R.M., and M.S. performed the measurements at TOMCAT. P.V.-P., B.P, R.M., and S.L. performed the experiments at ID01.
V.G. and C.D. fabricated the 3D nanosamples for the CDI experiments at ID01.
The project was supervised by M.S. and H.N.C. 
P.V.-P. and B.P. wrote the bulk of the manuscript.
All authors contributed to the preparation of the manuscript.

\section*{Competing interests statement}
\label{sec:conflictInterest}
The authors declare no competing interests.

\section*{Data Availability}
The data that support the plots within this paper and other findings of this study are available from the corresponding author upon reasonable request.

\bibliographystyle{unsrt}
\bibliography{paper}{}



\end{document}